\documentclass[runningheads]{llncs}

\usepackage{graphicx,subcaption}
\usepackage{hyperref}
\usepackage[T1]{fontenc}

\usepackage{booktabs}

\usepackage{xcolor}
\definecolor{candyapplered}{rgb}{1.0, 0.03, 0.0}
\newcommand{\ok}[1]{\textcolor{candyapplered}{\textsf{OK:~~#1}}}

\begin{document}

\title{Age Appropriate Design: Assessment of TikTok, Twitch, and YouTube Kids}
\titlerunning{Age Appropriate Design}
%
\author{Virginia N. L. Franqueira, Jessica A. Annor and Ozgur Kafali}
\authorrunning{Franqueira et al.}

%
\institute{School of Computing\\
University of Kent, Canterbury, UK\\
\email{V.Franqueira@kent.ac.uk, annor.j@gmail.com,  R.O.Kafali@kent.ac.uk}}
\maketitle              
\begin{abstract}
The presence of children in the online world is increasing at a rapid pace. As children interact with services such as video sharing, live streaming, and gaming, a number of concerns arise regarding their security and privacy as well as their safety. To address such concerns, the UK's Information Commissioner's Office (ICO) sets out 15 criteria alongside a risk management process for developers of online services for children. We present an analysis of 15 ICO criteria for age appropriate design. More precisely, we investigate whether those criteria provide actionable requirements for developers and whether video sharing and live streaming platforms that are used by children of different age ranges (i.e., TikTok, Twitch and YouTube Kids) comply with them. Our findings regarding the ICO criteria suggest that some criteria such as \emph{age verification} and \emph{transparency} provide adequate guidance for assessment whereas other criteria such as \emph{parental controls}, \emph{reporting of inappropriate content}, and \emph{handling of sensitive data} need further clarification. Our findings regarding the platforms themselves suggest that they choose to implement the simplest form of \emph{self-declared age verification} with limited parental controls and plenty of opportunities.
\end{abstract}

\section{Introduction}

Children (defined by the Information Commissioner's Office -- \emph{ICO}, UK) as ``a person under the age of 18 years''~\cite[Page 89]{ico2020}) are increasingly relying on online services for entertainment (e.g. to play games, to consume multimedia), for education (e.g. to support schoolwork, to learn skills), and for socialising (e.g. to establish connections, to communicate and participate in virtual communities).

\emph{Ofcom} (UK) conducted a study about children's use of media (3--15 years old) between April and July 2019. The study~\cite{ofcom2020-attitude-annex} involved ``2,343 in-home interviews with parents of children aged 5--15 and with children aged 8--15 were conducted, along with 900 interviews with parents of children aged 3--4''. The findings of the study~\cite[Page 5]{ofcom2020-attitude} show a range of online activities prevalent for under 16 years old such as watching videos, having social media profiles, and playing games, starting from a very young age. The results of another survey conducted by the \emph{EU Kids Online Network} among 25,101 9--16 years old across 19 European countries (between Autumn 2017 and Summer 2019)~\cite{lse-survey-2020} mostly corroborates with those trends, despite country-specific differences. It empirically observed ``a substantial increase in both the proportion of smartphone-using children and the amount of internet use''~\cite[Page 6]{lse-survey-2020} across all countries, compared to their previous survey from 2010. 

Despite potential benefits, children's intense presence online and connectivity to the Internet raise a number of security and privacy as well as safety (physical and mental well-being) concerns, both from online services and smart toys. The risks include exposure of personal identifiable information~\cite{nelson2016}, spying, tracking or profiling~\cite{sasha2019}, vulnerabilities such as voice injection (particularly concerning in relation to younger children)~\cite{valente2017}, online  grooming~\cite{kloess2019} and emerging exposure to cyberhate~\cite{lse-hate-2020} and promotion of gambling via in-game purchases.

To tackle some of the aforementioned threats, ICO's \emph{Age Appropriate Design} (2020)~\cite{ico2020} sets a code of practice for online services -- designed for or used by children -- to evaluate and justify compliance with the Data Protection Act 2018~\cite{dpa2018} and the General Data Protection Regulation (GDPR)~\cite{GDPR2018}. Moreover, ICO's Data Protection Impact Assessment (DPIA) guidance provides a risk management process for developers of online services for children.

This paper serves two purposes. For policy makers, we aim to identify the common areas of concern in online platforms that are popular among children. For developers of such platforms, we aim to provide guidance on which features to focus on and improve to protect children online. Specifically, to understand the ICO regulation and its impact on developers of online services for children, we explore two research questions:
\begin{itemize}
\item RQ$_1$: Does the ICO's code of practice provide sufficient guidance for developers of online platforms to implement their services following an age appropriate design?

\item RQ$_2$: Do video sharing and live streaming platforms that are popular among children comply with the ICO's code of practice for age appropriate design?
\end{itemize}

Our first contribution (in addressing RQ$_1$) is an analysis of the ICO's 15 criteria for age appropriate design to identify \emph{actionable} requirements (or lack thereof) for developers of online services for children, and a comparison of the ICO's 15 criteria with Ofcom's findings to identify any lacking issues that need to be addressed. Our second contribution (in addressing RQ$_2$) is an analysis of three online platforms that are popular among children: one video sharing platform (\emph{TikTok}), one live streaming platform (\emph{Twitch}), and a platform that is exclusively designed for children (\emph{YouTube Kids}). Through this analysis, we identify how the platforms comply with the ICO's 15 criteria for age appropriate design. We investigate the official documents posted on the platforms' websites (e.g. privacy policy, help, FAQ) for the analysis.

The rest of the paper is organised as follows.
Section~\ref{sec-background} explains the necessary background.
Section~\ref{sec-method} describes our methodology.
Section~\ref{sec-case} presents our study of the selected online platforms.
Section~\ref{sec-discussion} positions our work in the literature and discusses its significance.
Section~\ref{sec-conclusion} concludes the paper with future directions.

\section{Background}
\label{sec-background}

Ofcom's report \cite{ofcom2020-attitude} revealed the following statistics about children's online presence:

\begin{itemize}
    \item \textit{Own a tablet:} 24\% (3-4 years old), 37\% (5-7 years old), 49\% (8-11 years old) and 59\% (12-15 years old).
    \item \textit{Own a smartphones:} 0\% (3-4 years old), 5\% (5-7 years old), 37\% (8-11 years old) and 83\% (12-15 years old).
    \item \textit{Play games online:} 17\% (3-4 years old), 35\% (5-7 years old), 66\% (8-11 years old) and 72\% (12-15 years old).
    \item \textit{Watch YouTube:} 51\% for 8 hours 6 minutes a week (3-4 years old), 64\% for 8 hours 36 minutes a week (5-7 years old), 74\% for 10 hours a week (8-11 years old) and 89\% for 11 hours a week (12-15 years old).
    \item \textit{Have social media profile:} 1\% (3-4 years old), 4\% (5-7 years old), 21\% (8-11 years old) and 71\% (12-15 years old).
    \item \textit{Emerging online services:} newer platforms are becoming more popular, compared to the previous year survey, i.e. TikTok's use increased 13\% among 12-15 years old, and Twitch's use increased 5\%.
\end{itemize}

Having such a prevalent online presence of children brings forward the need for stringent control to ensure that human rights and well-being of children are enforced. For this purpose, regulations and legislation have been enacted prior to ICO's comprehensive set of requirements for online service providers.

The \emph{US Children's Online Privacy Protection Rule} (COPPA) \cite{coppa} ``sets forth a framework of fair information practices governing the collection, access to, and use of personal information by website directed to children [under the age of 13]. The Act does not apply to general audience websites; however, operators of such sites, who have specific sections for children or actual knowledge of children using their site, must follow the COPPA regulations. Also, COPPA applies to foreign websites that are directed at US children'' \cite{epic}.

The \emph{EU Audiovisual Media Services Directive} \cite{AVMSD} sets September 2020 as the deadline for implementation: ``Article 28b encompasses a series of duties of so-called video sharing platforms (VSPs) concerning the prevention and moderation of content \dots National authorities (mainly independent media regulatory bodies) are given the responsibility of verifying that VSPs have adopted \emph{appropriate measures} to properly deal with the types of content mentioned above (alongside other undesirable content). This includes the guarantee that platforms properly revise and enforce their ToS [Terms of Service]; have appropriate flagging, reporting, and declaring functionalities; implement age verification or rating and control systems; establish and operate transparent, easy-to-use and effective procedures to resolve users’ complaints; and provide media literacy tools'' \cite{barata2020}.

\section{Methodology}
\label{sec-method}

To answer RQ$_1$, we analyse the 15 criteria set out by the ICO regulation and identify whether actionable requirements for developers of online services for children are provided with clear guidance. Moreover, we compare ICO's 15 criteria with the findings of Ofcom's report to identify whether there are significant issues that need to be addressed in the ICO regulation.

To answer RQ$_2$, we investigate three video sharing and live streaming platforms. We analyse the official documents posted on their websites (e.g. privacy policy, help, FAQ) to determine whether they comply with ICO's 15 criteria. We removed ``DPIA'' from the criteria because it is part of our methodology for performing the analysis of the platforms. We also removed ``connected toys and devices'' from the criteria because the platforms we investigate are only accessible via the web or mobile devices. In cases where documentation lacked clarity, we created accounts within the platforms to test some of the features. We select three online platforms, TikTok, Twitch, and YouTube Kids, due to their popularity among children~\cite{ofcom2020-attitude}. To perform our analysis of the platforms, we adopt DPIA's steps, in particular Steps 2 and 4 regarding data processing. Step 2 -- ``Describe the processing'' asks the online service providers to describe the scope and purpose of their data collection, storage, and usage. Step 4 -- ``Assess necessity and proportionality'' asks the online service providers to describe their compliance with the 15 ICO criteria and additional issues regarding ethics and fairness, such as avoiding bias in AI algorithms. One author led the analysis of the platforms and all three authors discussed and validated the findings.

\section{Case Study}
\label{sec-case}

\subsection{RQ$_1$ -- ICO Requirements}

The ultimate aim of the ICO regulation is ``to ensure that online services likely to be accessed by children are appropriate for their use and meet their development needs''~\cite[Page 32]{ico2020}. Our investigation into the ICO criteria suggest that whereas some criteria such as age verification and transparency of privacy policies are clear, other criteria such as the extent to which parental controls and reporting of inappropriate content should be implemented and handling of sensitive data (e.g. children's location) need further detail. In particular, we report the following:

\begin{itemize}
\item ``Transparency'', ``Default settings'', ``Profiling'', and ``Nudge techniques'' provide clear instructions and examples of intuitive privacy notices and choices.

\item ``Age appropriate application'' covers options for age verification including automatic classification of age ranges using AI. However, it does not provide requirements for ensuring restricted access to content appropriate for age groups. It also overlaps with criteria such as ``Data minimisation''.

\item ``Best interests of the child'' and ``Data protection impact assessments'' cover a wide range of risks and commercial exploitation of children. However, it does not address specific concerns mentioned in the Ofcom report~\cite{ofcom2020-attitude} such as the promotion of in-game purchasing and \emph{loots} in video streaming platforms.

\item ``Geolocation'' recommends that the application should warn children whenever their location is in use and ask them to discuss this with an adult, but it does not recommend any monitoring options for parents.

\item Recommendation on ``Parental controls'' are only provided by the ICO if the application has such controls. The inclusion of parental controls is not a strict requirement. Moreover, the choices available to parents under ``Data minimisation'' and ``Data sharing'' are not clear.
\end{itemize}

\subsection{RQ$_2$ -- Online Platform Compliance}

We explain each platform in detail with references to their policies and interfaces. For each platform, we first start with a brief description of its online services and then investigate how it covers the ICO criteria.

\subsubsection{TikTok.} Anyone can access TikTok without an account -- search for videos, download videos (as long as allowed by their creators), view video comments, share videos, and report videos. Some limitations of the application for users without an account include the inability to like videos, follow profiles, comment on videos, record duet videos, and add videos to their list of favourites. For the analysis of the ICO criteria, we use the documentation available on the TikTok website with a focus on EEA and Switzerland policies\footnote{\url{https://www.tiktok.com/legal/privacy-policy?lang=en\#privacy-eea}\\ 
\url{https://www.tiktok.com/community-guidelines?lang=en}\\
\url{https://www.tiktok.com/legal/terms-of-use}}. 

``Best interest of the child'': Regarding the children's digital well-being, parents can restrict how much time their children spend on the app (Settings $\rightarrow$ Digital Wellbeing $\rightarrow$ Screen Time Management) -- 40 minutes, 60 minutes, 90 minutes or 120 minutes.

``Age appropriate application'': Individuals younger than 13 years old are not allowed to register for a TikTok account, according to their Terms of Service~\cite{tiktok-terms-of-use}. This is enforced by selecting the user's date of birth using the interface. If the calculated age is below 13, the process is halted, and the user does not have a second chance unless the TikTok application is uninstalled and re-installed. 

``Transparency'': TikTok is transparent regarding some of the features it provides. On the one hand, the user is given bite-sized and easy to understand explanation on why it uses ads and the implications of accepting personalised ads, together with a link with further information. It makes it clear that (not personalised) ads will be shown regardless. On the other hand, at the point of signing up, users are not informed of certain clauses in the privacy policy such as the following which gives unrestricted rights to content posted in the platform: ``\emph{by submitting User Content via the Services, you hereby grant us an unconditional irrevocable, non-exclusive, royalty-free, fully transferable, perpetual worldwide licence to use, modify, adapt, reproduce, make derivative works of, publish and/or transmit, and/or distribute and to authorise others users of the Services and other third-parties to view, access, use, download, modify, adapt, reproduce, make derivative works of, publish and/or transmit your User Content in any format and on any platform, either now known or hereinafter invented}''~\cite{tiktok-priv-policy}.

``Detrimental use of data'': Users can use \emph{coins} to promote their favourite TikTok stars. All accounts have associated wallets with a coins balance (recharges available, at the time of writing, ranged from 65 coins for $\pounds$0.99 up to 6607 coins for $\pounds$99.99; note that only account holders who declared to be 18 or above can buy coins and, for that, they can use Apple iTunes or Google Play accounts~\cite{tiktok-priv-policy}) and a gift count (which keeps track of the gifts received by other users).

``Policies and community standards'': TikTok's Community Guidelines~\cite{tiktok-community} set best practices of the type ``do not post'' for content related to a number of subjects, e.g. hate speech, harassment and bullying, minor safety, and adult nudity and sexual activities. Content deemed inappropriate by users can be reported. Consequences include content removal, account bans or suspensions, and legal action~\cite{tiktok-community}.

\begin{figure}[htbp]
	\begin{subfigure}{0.5\linewidth}
		\centering
		\includegraphics[width=1\textwidth]{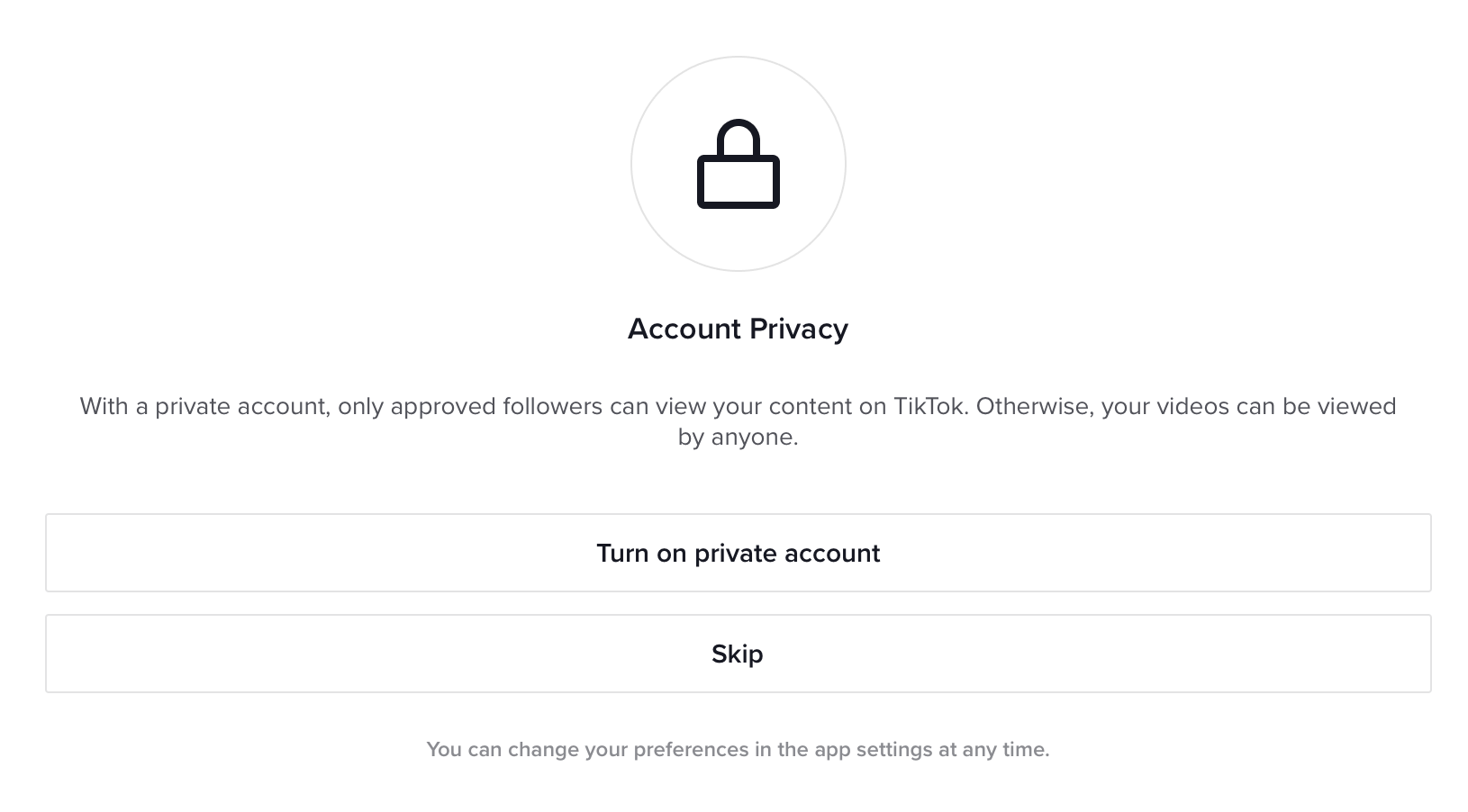}
		\caption{Turning on private account\label{fig:tiktok-private-1}}
	\end{subfigure}\hfill
	\begin{subfigure}{0.5\linewidth}
		\centering
		\includegraphics[width=1\textwidth]{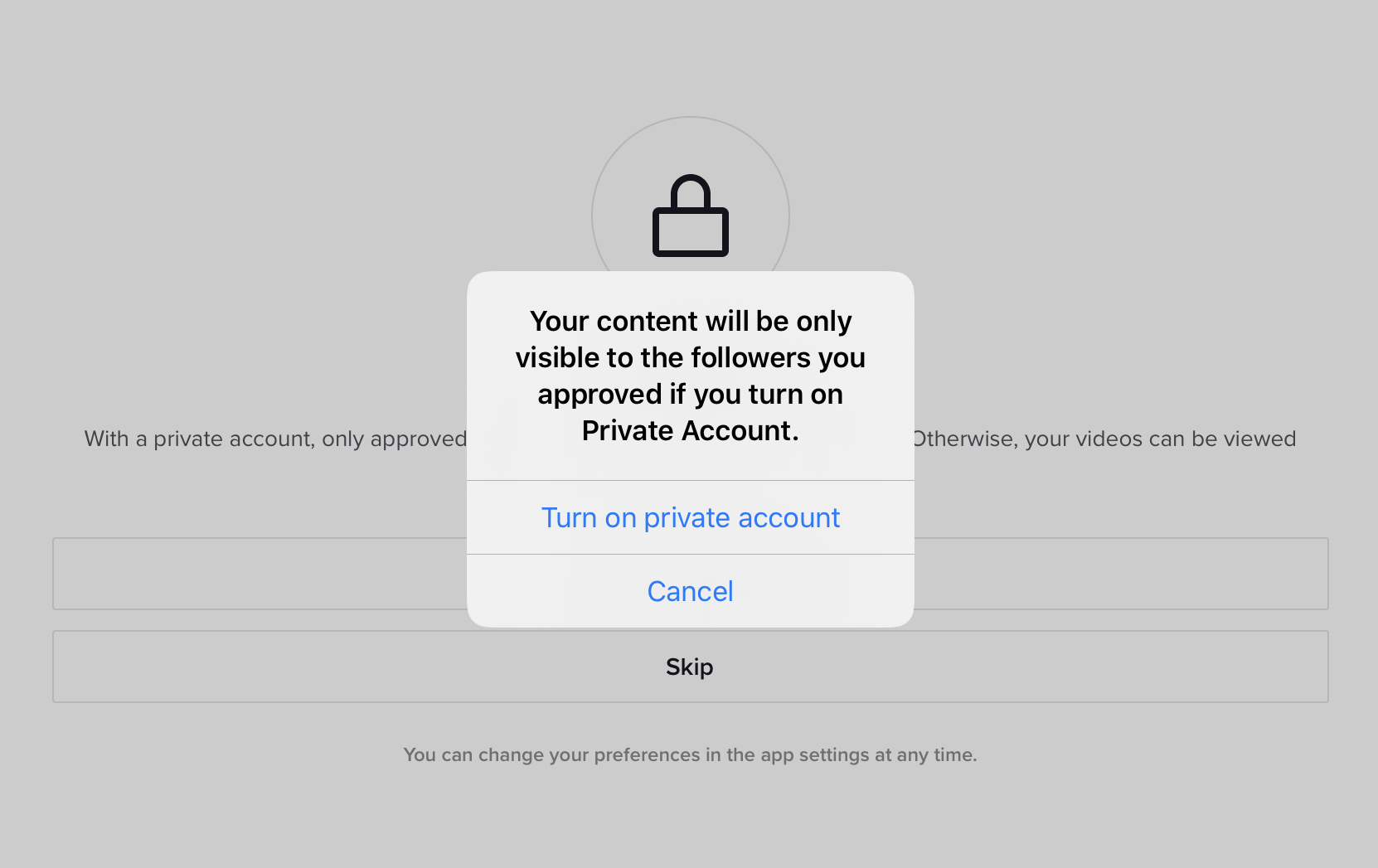}
		\caption{Warning before completion\label{fig:tiktok-private-2}}
	\end{subfigure}
	\caption{Steps to change a TikTok account from the default ``public'' to ``private''.}
	\label{fig:tiktok-private}
\end{figure}

``Default Settings'': By default, new accounts are \emph{public}, unless turned ``private'' (Fig.~\ref{fig:tiktok-private-1}). Before turning an account to private, a warning appears to the user: \emph{Your content will be only visible to the followers you approved if you turn on Private Account} (Fig.~\ref{fig:tiktok-private-2}). By default, only \emph{friends} of a TikTok user can send messages to that user. Moreover, TikTok selects a number of user accounts to suggest to other users (i.e. to strangers from outside their circle of connected accounts), prompting them to send a friend request. When users create their account, they are asked if they want to allow TikTok to show them personalised ads. However, the response is biased in favour of the option \emph{Accept}. 
Multiple TikTop users can log in on a same device, which makes privacy settings personal to individual account holders (e.g. parent vs child).

``Data minimisation'': Their privacy policy~\cite{tiktok-priv-policy} states that TikTok automatically collects device information, e.g. device type and operating system. It also collects \emph{keystroke patterns}, for which the purpose is not specified.

``Data sharing'': As mentioned in their privacy policy, TikTok may share data with law enforcement or with third-parties. A user's device ID may be made public to advertisers. TikTok may also share data with analytic, service and payment providers, as well as its business partners. When a user signs up using Facebook, Google or Twitter, their public profile is shared with these companies.

``Geolocation'': TikTok collects the users' GPS by default and does not provide the user the option to turn off GPS via the application~\cite{tiktok-priv-policy}. The user needs to explicitly switch-off location via mobile device settings.

``Parental controls'': TikTok provides two main features for parental controls, \emph{Restricted Mode} and \emph{Family Pairing}. When turning on restricted mode on their child’s device for the first time, parents are prompted to enter and confirm a password (4-digit PIN). In restricted mode, the child is not able to access content flagged by parents. Family pairing allows parents to create a family group to monitor and manage their children's use of the application. To do so, the parent first creates an account and navigates to Settings $\rightarrow$ Family Pairing, where a unique QR-code is generated. The parent then -- using their child’s device -- navigates to Settings $\rightarrow$ Family Pairing and declares the account as a \emph{teen account}. Upon doing so, the parent is prompted to scan the QR code. Once parent and child accounts are linked, the parent' settings would apply to the child's account. If the child tries to unlink the account, the parent gets a notification.

``Profiling'': TikTok's privacy policy makes it clear that the application (including unnamed third-parties) uses cookies to collect information and analyse usage. With web beacons (i.e. pixel-size data embedded in images~\cite{tiktok-priv-policy}), they are able to track fine-grained usage information, e.g. timestamp and description of pages visited, and ``information from your computer or device''~\cite{tiktok-priv-policy}. The application does not provide a way to refuse or disable cookies; though the policy refers to disabling mechanisms from browsers and mobile device features.

``Nudge techniques'': TikTok displays a warning when a user attempts to change their account from private to public: ``Your content will be only visible to the followers you approved if you turn on Private Account''. 

``Online tools'': TikTok users are not automatically notified when their accounts are logged into from other devices. Instead, they have to manually navigate to -- Settings $\rightarrow$ Manage my account $\rightarrow$ Security -- frequently to monitor access to their account. A list of alerts show all devices that have logged into the specific TikTok account for the last 7 days with login timestamp and timezone. The user can then select a device that is unfamiliar and click \emph{Secure my account}.
Users can contact \emph{feedback@tiktok.com} to have an account deleted. There are no explicit indicators regarding time it will take or assurance regarding how comprehensively data will be deleted from all third-parties. Regarding \emph{duet} videos, TikTok provides options for the original creators to control who can create a duet video with them: ``everyone'' (anyone with a TikTok account), ``friends'', or ``no one''~\cite{tiktok-duet}. However, TikTok documentation does not make it clear whether there is a mechanism in place to delete (or request the deletion of) an unwelcome duet of ``my videos'', unless by direct interaction with the duet's creator.


\subsubsection{Twitch.} Anyone can access Twitch's content (organised in channels or categories), and perform a number of actions such as search or browse for content or for users, watch video-on-demand (previously streamed) or live video streams, discover ``recommended'' content, read stream chats, share a video/stream, and view user profiles. Signing up for an account allows access to additional features such as adding comments to stream chats, privately sending messages to other users (also called \emph{whispers}), making purchases, and giving and receiving gifts. For the analysis of the ICO criteria, we use the documentation available on the Twitch website\footnote{\url{https://www.twitch.tv/p/en-gb/legal/terms-of-service/}\\
\url{https://www.twitch.tv/p/en-gb/legal/privacy-notice/}\\
\url{https://www.twitch.tv/p/en-gb/legal/community-guidelines/}\\
}.

``Best interest of the child'': Although Twitch's most prominent use is to stream video games, channels under the category \emph{just chatting} is gaining popularity, as indicated by recent statistics\footnote{https://sullygnome.com/games/365}. Some content in this category may potentially affect children's mental well-being, and development (e.g. sensual behaviour and traits shown by streamers). The application has some characteristics leading to commercial exploitation. For example, it is possible to turn-off third-party ads but not Amazon ads, channel streamers accept ``donation'' and may add an Amazon wishlist to their page, and Amazon account holders (e.g. ``prime student'' or ``Twitch prime''\footnote{https://www.twitch.tv/p/en-gb/students/}) have privileges in the application.

``Age appropriate application'': The input of a username, password, date of birth and email address are necessary information to create a Twitch account. Twitch sends a verification code via email that is required to activate the account. Section 12 of Twitch's \emph{Privacy Notice}~\cite{twitch-priv-policy} regards ``Children's Privacy'': \emph{``If you are under 13 years of age, then please do not use or access the Twitch services at any time or in any manner}''.

``Transparency'': Twitch documentation is comprehensive. Particularly helpful is their \emph{Privacy Choices}~\cite{twitch-priv-choices} document which provides tables with two columns of the type: ``If you want to \dots'' $\rightarrow$ ``Do this \dots''. For example: 
\begin{itemize}
    \item Disable or deactivate your Twitch user account $\rightarrow$ Log into your Twitch account and go to your Settings page
    \item Block whispers from strangers $\rightarrow$ Go to the Security and Privacy tab in your Settings
    \item Opt out of marketing emails from Twitch $\rightarrow$ Use the ``unsubscribe'' link included in each email
    Restrict the way that we process and disclose specific personal information about you $\rightarrow$ Submit your request here
\end{itemize}

\begin{figure}[hbt]
	\begin{subfigure}{0.5\linewidth}
			\centering
		\includegraphics[width=0.95\textwidth]{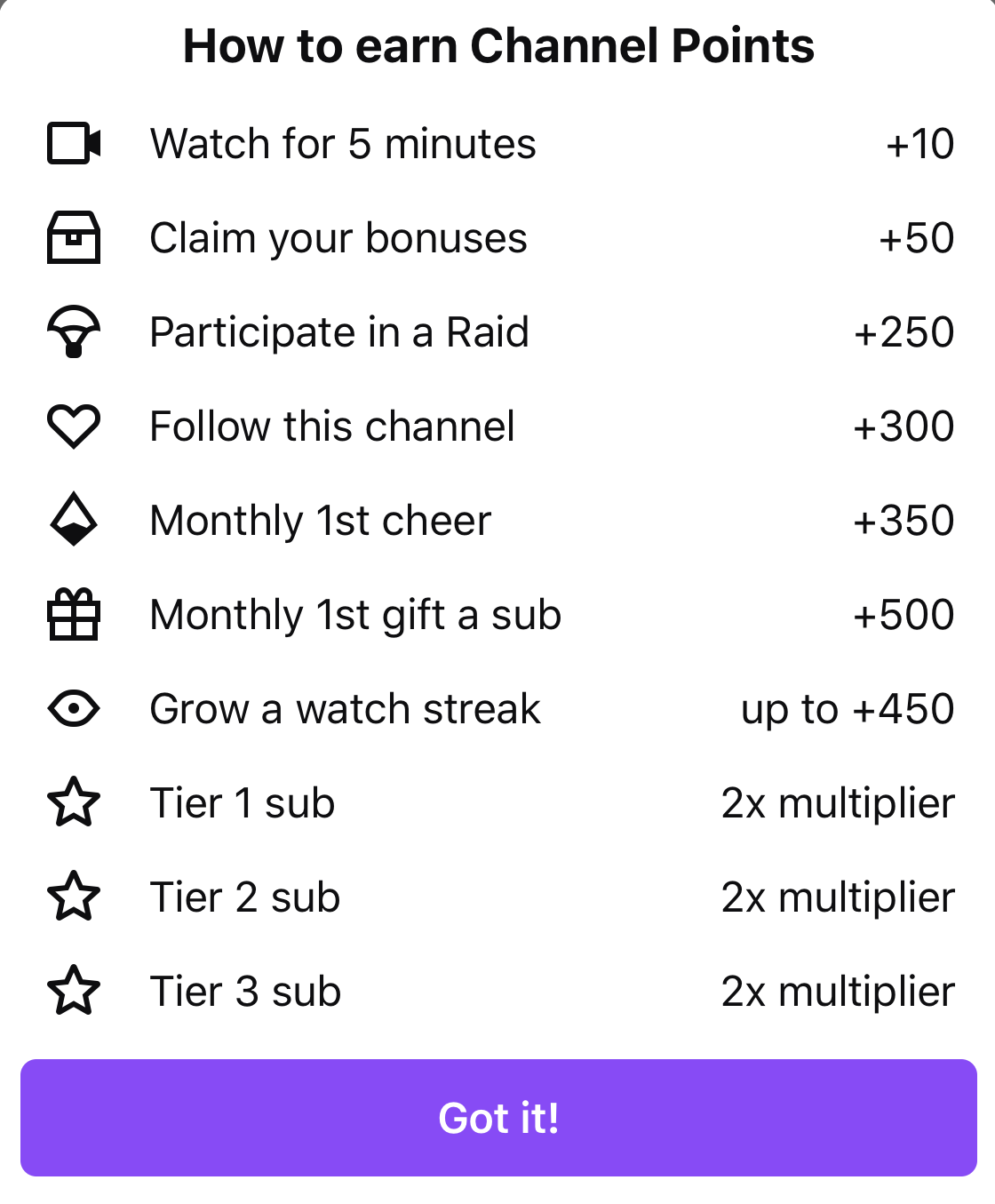}
		\caption{Twitch: How to earn points' table.\label{fig:twitch-points}}
	\end{subfigure}\hfill
	\begin{subfigure}{0.5\linewidth}
			\centering
		\includegraphics[width=1.1\textwidth]{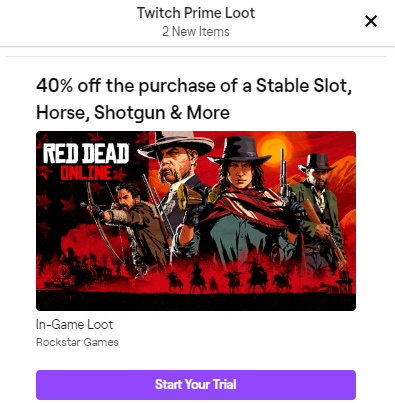}
		\caption{Twitch Prime Loot feature.\label{fig:twitch-loot}}
	\end{subfigure}\hfill	
	\caption{Two features of Twitch that might affect mental health well-being.}
	\label{fig:twitch-features}
\end{figure}

``Detrimental use of data'': Twitch encourages users to compete and spend money with gambling-like features. One of its features is to accumulate points via a range of activities, as shown in Fig.~\ref{fig:twitch-points}. Some of these activities require payment which can be done via debit or credit card, and via mobile payment. One way of earning points is to participate in \emph{Raids}. This happens when, at the end of a video streaming session in a channel, the streamer announces to viewers a raid at another channel, which may be unknown to them. Individual viewers are not obliged to participate in the announced raid. However, excitement and ``group behaviour'' may encourage participation. Users can also buy \emph{Sub} tokens that Twitch distributes among a streamer's subscribers. The cost of one Sub token is equivalent to $\pounds$5.99 (June 2020). Twitch has an algorithm that determines the recipients of the Sub tokens. As such, 100 random Twitch users are rewarded with a \emph{gift}, i.e. a free subscription to a channel. Another feature of Twitch is the \emph{Community Gift Chest} that streamers can enable for their channel. Twitch account holders with a linked Amazon Prime subscription can contribute with gifts to such chests using the called \emph{Twitch Prime Loot} feature, exemplified in Fig.~\ref{fig:twitch-loot}; the process stops when the streamer decides to distribute the accumulated gifts in their chest~\cite{twitch-loot}. All viewers of the channel who comply with eligibility criteria (i.e. not Prime members, had been watching channel with Twitch account, and first-time receivers of gifts) may be randomly selected. 

``Policies and community standards'': Parents and legal guardians are instructed~\cite{twitch-priv-policy} to report to \emph{privacy@twitch.tv} upon detection of under 13 years old children using Twitch. If the child holds an account, Twitch promises to permanently delete it, as well as any personal information associated with that account. Twitch's \emph{Community Guidelines}~\cite{twitch-community} cover a number of topics, e.g. ``Hateful Conduct and Harassment'', ``Unauthorized Sharing of Private Information'', and ``Nudity, Pornography, and Other Sexual Content''. The latter is further elaborated on a \emph{Sexual Content} guideline~\cite{twitch-nudity-policy} with the following explanation: ``\emph{While we understand that some nudity or sexual content might be intended for educational, scientific, artistic, newsworthy, or academic purposes, we restrict this content due to the diversity in age and cultural backgrounds of our global community''}~\cite{twitch-nudity-policy}. Violations of policy and guidelines may lead to ``removal of content, a strike on the account, and/or suspension of account(s)''~\cite{twitch-community}.

\begin{figure}[hbt]
	\begin{subfigure}{0.5\linewidth}
			\centering
		\includegraphics[width=0.9\textwidth]{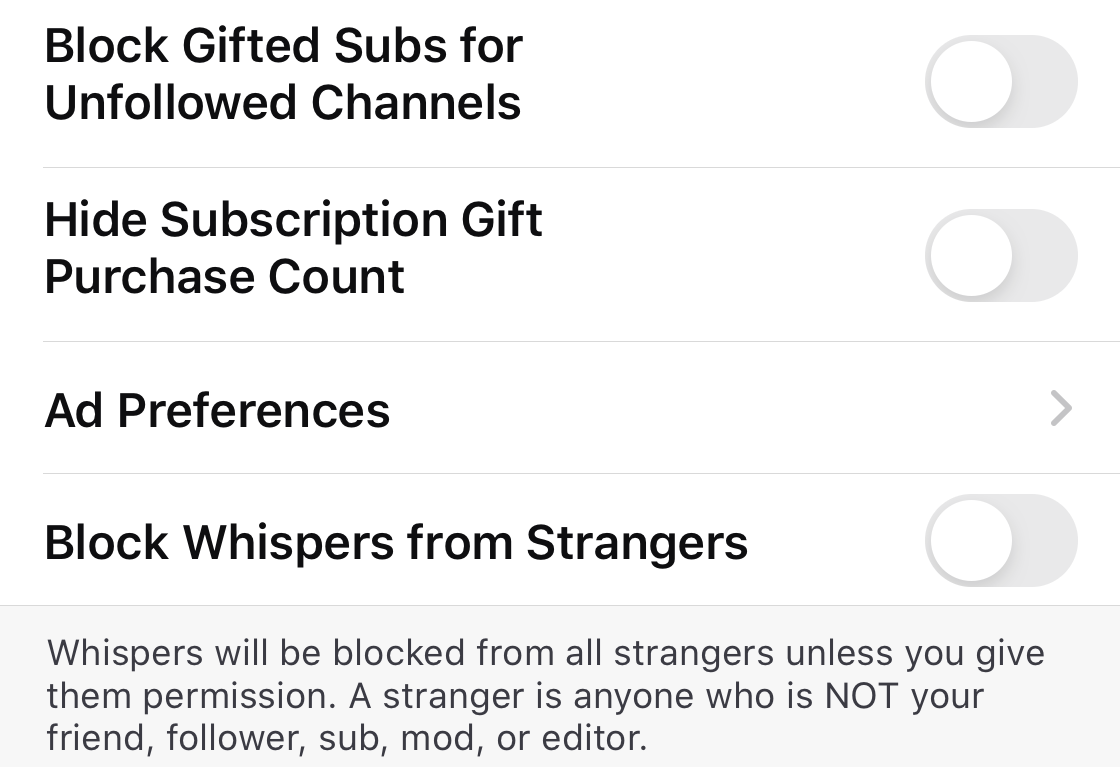}
		\caption{Privacy settings turned off.\label{fig:twitch-defaults}}
	\end{subfigure}\hfill
	\begin{subfigure}{0.5\linewidth}
			\centering
		\includegraphics[width=0.9\textwidth]{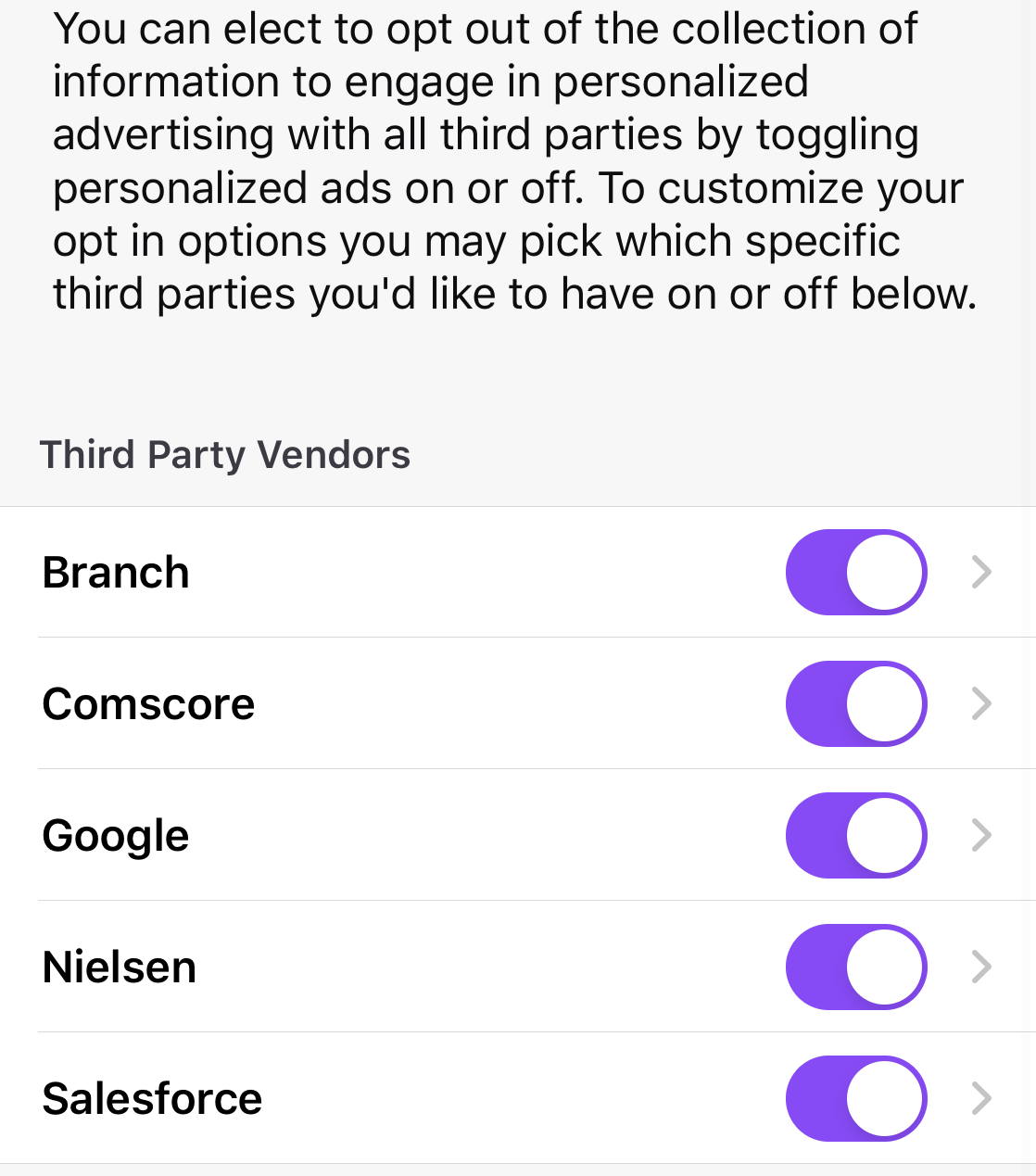}
		\caption{Third-party advertisement turned on.\label{fig:twitch-ads}}
	\end{subfigure}\hfill	
	\caption{Examples of Twitch default settings.}
	\label{fig:twitch-settings}
\end{figure}

``Default Settings'': By default, accounts are set to ``Allow all raids”. All notifications (i.e. Twitch app notification, email notifications and mobile phone notifications) are turned on by default. Under the \emph{Privacy} tab in Settings, the following is turned off by default: Block Gifted Subs for Unfollowed Channels, Hide Subscription Gift Purchase Count, and Block Whispers from Strangers (Fig~\ref{fig:twitch-defaults}). Third-party advertisements and cookies are turned on by default, unless ad preferences are tuned (as shown in Fig.~\ref{fig:twitch-ads}); one caveat is the impossibility to turn off Amazon ads.

``Data minimisation'': Twitch collects a vast amount of data while providing their services; this includes three categories of data. First, data provided by the users themselves, e.g. ``name, voice and image (in any content you upload), Twitch username, email address, postal mailing address, telephone number, credit card number, and billing information''~\cite{twitch-priv-policy}. Second, data collected automatically by cookies and other technologies (e.g, web beacons) including ``IP addresses, a unique user ID, device and browser types and identifiers, referring and exit page addresses, software and system type''. Usage data is also automatically collected by Twitch and by third-parties providers of personalised ads, listed in Fig.~\ref{fig:twitch-ads}. Third, data obtained from a range of third-parties such as ``from advertisers, games or services you use, or social media networks (such as Facebook) for which you have approved our access''~\cite{twitch-priv-policy}. Twitch \emph{Cookies Policy}~\cite{twitch-cookies} mentions explicitly that they do not   control cookies used by third-parties.

``Data sharing'': Twitch shares data with a number of third-parties; five of them are listed in Fig. 5b: Branch, Comscore, Google, Nielson and Salesforce. This can, in principle, be controlled by users although data sharing is allowed by default. Being an Amazon company, Twitch data is automatically shared this way without any user control. Moreover, Twitch users typically advertise their presence in social media and other services and, despite any consent given by the channel streamer, it is not clear what information is automatically shared with those parties about other users interacting with them. 

``Geolocation'': Twitch documentation does not make it clear whether GPS data is automatically collected from the device been used to connect to their services. However, user provided data at time of account creation includes postal address. Information from September 2018\footnote{https://discuss.dev.twitch.tv/t/get-geolocation-of-stream-viewers/17647} indicates that their API for developers did not provide GPS information at the time, and IP address was the best approximation possible.

``Parental controls'': Twitch does not provide any form of parental control. 

``Profiling'': Usage data and cookies are used for profiling purposes to recommend content, offers, and advertising. Twitch refers users to \emph{Security \& Privacy} settings of mobile devices.

\begin{figure}[h]
		\centering
		\includegraphics[width=.8\textwidth]{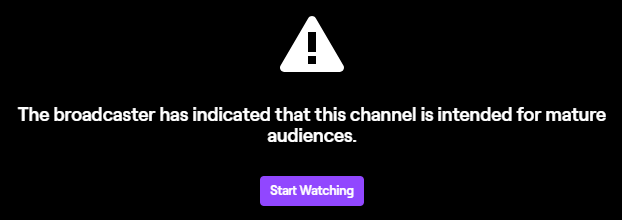}
		\caption{Twitch's content warning; can be bypassed with the purple option.}
		\label{fig:twitch-restricted}
\end{figure}

``Nudge techniques'': When a user tries to access a channel intended for a mature audience, the warning shown in Fig.~\ref{fig:twitch-restricted} appears.

\begin{figure}[htb]
	\begin{subfigure}{0.5\linewidth}
			\centering
		\includegraphics[width=0.62\textwidth]{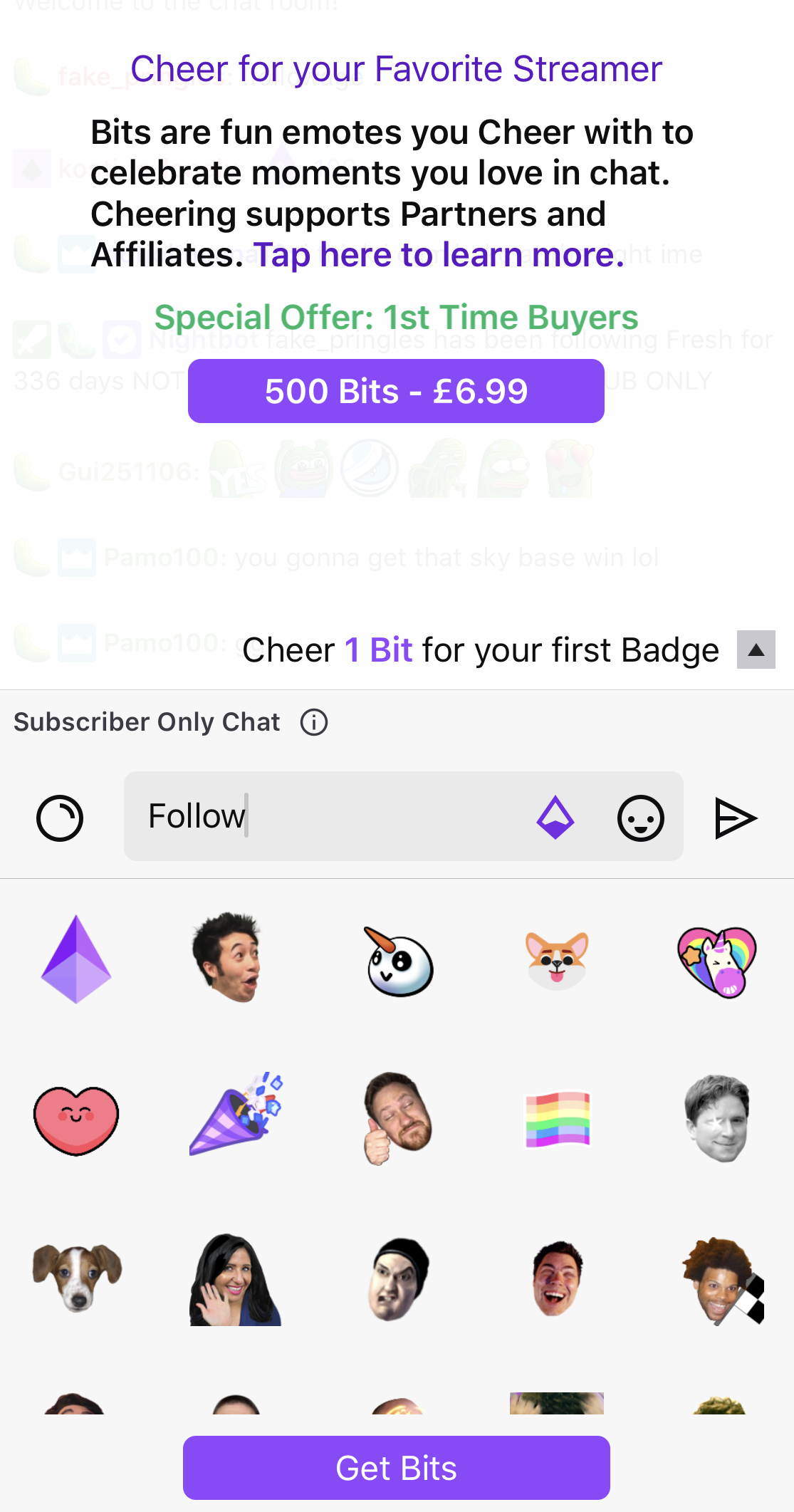}
		\caption{Nudge: \emph{Bits} as give-away reward.\label{fig:twitch-bits}}
	\end{subfigure}\hfill
	\begin{subfigure}{0.5\linewidth}
			\centering
		\includegraphics[width=0.62\textwidth]{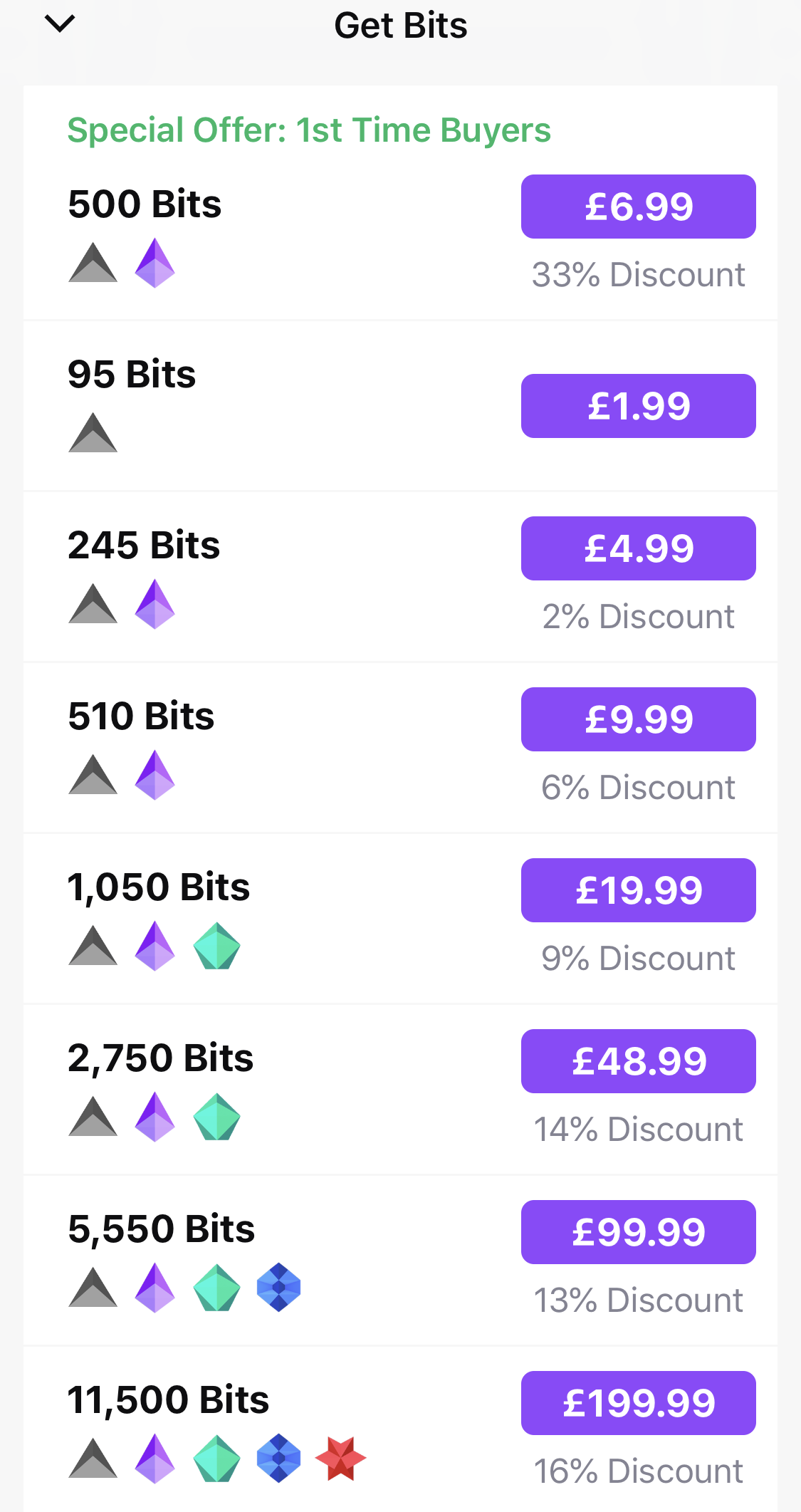}
		\caption{Nudge: buy pack of  \emph{Bits}.\label{fig:twitch-bits-price}}
	\end{subfigure}\hfill	
	\caption{Example of nudge techniques used by Twitch.}
	\label{fig:twitch-nudges}
\end{figure}

When registered users try to add comments to a stream chat on a specific channel, they are nudged to support their favourite streamers with \emph{Bits}. Bits are special colourful stickers that can be shared in chats, which financially benefits the streamers. They cost money and have various prices for each package (see Fig.~\ref{fig:twitch-nudges}). Wording  such as ``special price'', ``offer'', and ``promotion'' encourage spending by users or their parents.

``Online tools'': Twitch allows streamers to set an AI-based moderation tool, called ``AutoMod'' for their channel~\cite{twitch-moderation}. The tool automatically withholds potentially inappropriate comments in stream chats in terms of ``identity, sexual language, aggressive speech, and profanity''~\cite{twitch-moderation} to be manually approved or denied by the streamer's moderators. 
Twitch allows users to request deletion of account and associated data~\cite{twitch-priv-policy}. Users from EEA can make further privacy-related requests such as rectify or transfer personal information, and restrict processing or disclosure of personal data~\cite{twitch-priv-choices}.

\subsubsection{YouTube Kids.} YouTube is deemed appropriate for individuals aged 13+, and its child-friendly version -- YouTube Kids -- caters for children under 13 years of age. For the analysis of the ICO criteria, we use the documentation available on the YouTube Kids website\footnote{\url{https://kids.youtube.com/t/privacynotice}\\
\url{https://support.google.com/youtubekids/\#topic=6130504}\\
\url{https://kids.youtube.com/t/terms}\\
}.

\begin{figure}[htbp]
	\begin{subfigure}{0.5\linewidth}
		\centering
	    \includegraphics[width=1.1\textwidth]{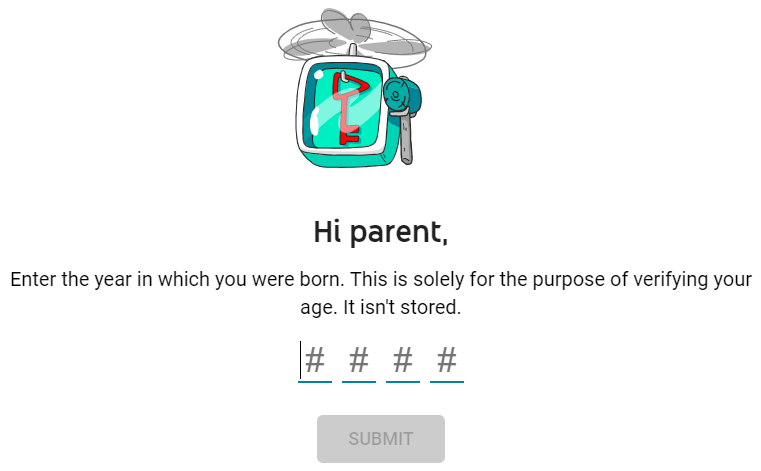}
		\caption{Input of parent's year of birth\label{fig:youtube-kids-1}}
	\end{subfigure}\hfill
	\begin{subfigure}{0.5\linewidth}
		\centering
		\includegraphics[width=0.8\textwidth]{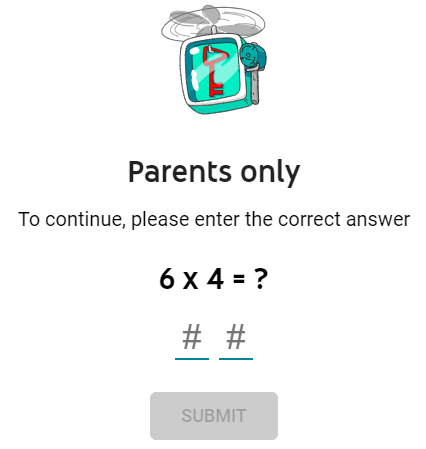}
		\caption{Input of multiplication result\label{fig:youtube-kids-2}}
	\end{subfigure}
	\caption{YouTube Kids (Web version): Parents' age verification process.}
	\label{fig:youtube-kids-age}
\end{figure}

Upon accessing YouTube Kids for the first time, a selection has to be made between \emph{I'm a child}, which prompts ``Ask a parent to set up YouTube Kids'', and \emph{I'm a parent}, which prompts self-declaration of the parent's year of birth (Fig.~\ref{fig:youtube-kids-1}). The parent is then shown an animation instructing to sign in with a parent's account which allows blocking videos and channels for their associated kid's account.

``Best interest of the child'': Parents can set different time limits (up to 60 minutes) on the use of the application for each of their associated children profiles~\cite{youtube-kids-parental-guide}. When the time set is up, the content on the kid's device is interrupted. This can be bypassed by input of parents' passcode (4-digits PIN, when explicitly configured), or by successfully answering a primary-school multiplication question (Fig~\ref{fig:youtube-kids-2}), or by leaving the application and reloading it.

``Age appropriate application'': In terms of \emph{age verification}, to create a child's profile, parents need to provide their year of birth, as shown in Fig.~\ref{fig:youtube-kids-1}. No age verification is enforced apart from a primary-school level multiplication question (Fig.~\ref{fig:youtube-kids-2}). It is possible to attempt this step multiple times by leaving the application and reloading it. Parents are then prompted to sign in with a Google account (in some countries~\cite{youtube-kids-priv-policy}) to customise profiles for their children. In terms of \emph{age classification}, a parent account can be used to create up to 8 child profiles. As part of profile setting, an age range has to be informed; options are:
``preschool'' (4 or under), ``younger'' (5--7) and ``older'' (8--12). 

``Transparency'': Before a parent can setup a YouTube Kids profile for their child, a \emph{privacy notice} has to be confirmed via an \emph{I Agree} button; the notice is short, but it contains technical jargon as shown in the following policy extract: 
``\emph{Information we collect~\cite{youtube-kids-priv-policy} \dots YouTube Kids collects information based on your child’s use of the app, like when they watch a video; This information includes: device type and settings, such as hardware model and operating system version; log information, including details of how our service is used, device event information and the device’s Internet protocol (IP) address; \dots''}.

YouTube Kids also has a privacy notice for children~\cite{youtube-kids-notice-children}, which is meant to be read through by kids with assistance from their parents. The privacy notice for children basically contains the same headings as the privacy notice for parents but without technical jargon to make it more accessible as shown in the following policy extract: ``\emph{What information do we collect?~\cite{youtube-kids-notice-children} We collect information that a parent may give us, like your name and month and year of birth, and we collect information as you use YouTube Kids. This may include things like your search and watch history and information about your device.}''

``Detrimental use of data'': YouTube Kids'  
privacy policy states that ``We use unique identifiers to provide contextual advertising, including ad frequency capping. The app does not allow interest-based advertising or remarketing''~\cite{youtube-kids-priv-policy}. On the other hand, YouTube Kids' Terms of Service~\cite{youtubekids-terms} allow the uploading of content promoting business or artistic enterprise. In practice, videos advertising merchandise in a subtle way, e.g. the \emph{GirlsTToyZZ} channel and videos of small children \emph{visiting} toy shops appear, e.g. in ``preschool'' kids profile.

``Policies and community standards'': An account can be used to create content, i.e. upload videos to YouTube Kids through \emph{YouTube Studio}, declaring them as child-friendly. A field-guide with practical tips for creation of content specific to YouTube Kids is available~\cite{youtube-create-content}.
Although, there is no \emph{community standard} specific to YouTube Kids, just an overarching one for YouTube~\cite{youtube-community}, it points to material specific to \emph{child safety} on YouTube~\cite{youtube-child-safety-1,youtube-child-safety-2}. Such material expands on content creation with some concrete examples.   

``Default Settings'': By default, YouTube Kids does not allow its users to comment on videos.

``Data minimisation'': According to YouTube Kids's Privacy Notice~\cite{youtube-kids-priv-policy}, no personal information is collected from kids directly. However, children may use voice-activated search (when search has been turned on by parents). The policy states that voice collected is immediately deleted after the search is processed. Children's profiles, filled by a parent account, contain name and month/year of birth. Parents can personalise their children's profiles with a display image, which are limited to a selection of animated characters.

``Data sharing'': YouTube Kids' Privacy Notice makes it explicit that information is shared with third-parties upon consent: \emph{``We may share individual user information with companies, organisations or individuals outside of Google when we have parental consent''}~\cite{youtube-kids-priv-policy}. Data is also shared for a number of reasons such as legal reasons, with partners for processing reasons, and to detect potential violation of the Terms of Services~\cite{youtubekids-terms}.  

``Geolocation'': The application collects information from the device~\cite{youtube-kids-priv-policy}. It is not clear which information precisely, and whether GPS data is included. It also collects the IP address associated with the device using the application which can give an indication of location, if proxies are not used.

``Parental controls'': The parental guide~\cite{youtube-kids-parental-guide} explicitly mentions that the YouTube Kids mobile app (as opposed to the Web version) is the only way for parents to approve, block or choose preferred content (collections, channels and videos) in relation to their children's profiles, increasing control.   For the preschool range, the application makes it clear that ``not all videos have been manually reviewed'', therefore, prompting the parent to flag it (``Our staff carefully reviews flagged content 24 hours a day, 7 days a week''~\cite{youtube-community}). At this range, parents are also prompted to make a decision regarding turning search on or off, therefore, allows preschool kids to search on their own.
A parent can also clear the watch and search history of their children's profiles, via Settings $\rightarrow$ Clear History. To access such settings, it is a matter of clicking on the lock icon and solve a primary-school multiplication question (Fig.~\ref{fig:youtube-kids-2}), therefore, easy to bypass. For an increasing security, a strong passcode or strong Google account password has to be configured, at the discretion of parents.

``Profiling'': Usage data (e.g. watch and search history) are used for profiling purposes to recommend content.  

``Nudge techniques'': When uploading content, a warning prompts the content creator to consider if there are children in the upload, emphasising that it may expose them to ``harm, exploitation, bullying'', and may violate laws.

``Online tools'': Parents can delete their children’s profiles. For that, the parent has to navigate to Settings, select the child's profile to delete, choose a three-dots option, and confirm deletion. Deleting a child’s profile deletes their watch history across devices. YouTube Kids does not provide any mechanism for monitoring login to a child's account from different devices.

\section{Discussion and Related Work}
\label{sec-discussion}

\begin{table}[h!]
\centering
\caption{Comparison of online platforms with regards to the ICO criteria.}
\begin{tabular}{p{2.5cm} | p{3cm} | p{3cm} | p{3cm}}
\toprule
\textbf{ICO Criteria} & \textbf{TikTok} & \textbf{Twitch} & \textbf{YouTube Kids}\\
\midrule
Best interest of the child & Limited screen time & Promotion of Amazon ads & Limited screen time; in-app ads\\[5pt] \midrule
Age appropriate application & Self-declared date of birth & Self-declared date of birth & Self-declared year of birth; multiplication test\\[5pt] \midrule
Transparency & Bite-sized explanations & Privacy choices document & Short explanations; no technical jargon for children\\[5pt] \midrule
Detrimental use of data & Incentive for in-app purchases & Gambling-like features & Contextual advertising; promotion of merchandise\\[5pt] \midrule
Policies and community standards & Content reporting & Content removal; account suspension & Content reporting\\[5pt] \midrule
Default Settings & Public accounts; friend-only messages; multiple accounts & Whispers from strangers blocked & Video comments disabled\\[5pt] \midrule
Data minimisation & Device properties; keystroke patterns & Personal information; device properties & No personal information\\[5pt] \midrule
Data sharing & Law enforcement; third-parties & Third-parties & Legal reasons; third-parties\\[5pt] \midrule
Geolocation & GPS enabled by default & IP addresses & IP addresses collected by default\\[5pt] \midrule
Parental controls & Restricted mode; family pairing & None & Restrict content; clear history\\[5pt] \midrule
Profiling & Cookies; targeted ads & Cookies; personalised ads; recommend content and offers & Recommend content; contextual advertising\\[5pt] \midrule
Nudge techniques & Public account warning & Channel warning; support streamers & Content upload warning\\[5pt] \midrule
Online tools & Delete account; monitor access from devices & Chat moderation & Delete profile; monitor watch and search history from devices\\[5pt]
\bottomrule
\end{tabular}
\label{tbl-checklist}
\end{table}

Table~\ref{tbl-checklist} summarises our findings of the three platforms against the ICO criteria. Across all three online platforms analysed, we observed the following:

\begin{itemize}
\item Individuals, regardless of their age, can access and view content provided by those platforms.

\item The weakest form of age verification, i.e. self-declared age via date of birth, is adopted by those platforms.

\item The platforms make a good attempt to present their privacy notices in a simple and understandable manner.

\item The content provided by the platforms (through their users as content creators) enable the promotion of in-app or in-game purchases, which is a concern for the well-being of children. Whereas this promotional behaviour might be unintentional in most situations, this could lead to gambling-like behaviour for children. Note that some countries including the UK are debating to ban loot boxes from games. The UK's \emph{The House of Lords Select Committee} mentioned in a recent report\footnote{\url{https://www.parliament.uk/business/committees/committees-a-z/lords-select/gambling-committee/news-parliament-2019/lords-gambling-report-published/}} that: ``The Government must act immediately to bring loot boxes within the remit of gambling legislation and regulation.''

\item The platforms have clear community guidelines for content creators and enable the reporting of potentially harmful content and accounts.

\item The platforms collect a variety of data through the use of cookies that allows profiling of children, and such data might be shared with third parties.

\item The platforms provide a limited number of warnings to its users, mainly regarding potentially harmful content, and often rely on the community or parents for reporting.
    
\end{itemize}

Apart from the common areas of concern, we highlight platform-specific remarks. YouTube Kids provides fewer features as part of their default settings (e.g. comments are disabled for videos), and a wider range of parental controls. This is not surprising as this is a platform exclusively for under 13 years old -- i.e. smaller children. TikTok provides the ability to create video duets, as long as allowed by the creator. This is appealing to raise popularity but the creator has no control about the type of duet. Twitch provides streamers with AI-based chat moderation tools to identify potentially harmful content posted by viewers. However, it depends on streamers to set their own moderators to manually \emph{accept} or \emph{deny} messages flagged automatically by the tool.

Analysis of compliance to laws and regulations has been object of intense research. Reyes et al.~\cite{reyes2018} studied compliance of popular free Android children's apps to COPPA. Violation to GDPR has also gained attention since its implementation in 2018; for instance, in relation to requirements of \emph{consent} for collection and processing of personal data~\cite{nouwens2020,utz2019}. Apthorpe et al.~\cite{apthorpe2019} took a different angle, and analysed the alignment of COPPA requirements to the perception of privacy from 195 parents in relation to smart toys. 

Extracting actionable security and privacy requirements from regulatory text, such as the ICO's age appropriate design criteria, is important for developers of online platforms for children. However, this is often a tedious manual task for security analysts. Complementary approaches based on formal logic, natural language processing (NLP), and crowdsourcing have been proposed to assist this process. Breaux et al.~\cite{Breaux+Anton-08:regulatory,massey2010} developed a systematic methodology for manually extraction of rules (e.g. obligations) that govern software systems, and presented a case study on the Health Insurance Portability and Accountability Act \cite{hipaa03privacy}. Hashmi \cite{Hashmi2015} proposed the extraction of legal norms from regulatory documents for compliance checking. Riaz et al. \cite{riaz14re} developed a machine learning tool for identifying security requirements from text, based on a set of context-specific requirement templates. Kafal{\i} et al. \cite{Kafali-17:Semaver} proposed a formal logic-based framework that semantically compares what happened in a security breach with what the regulatory text states. They developed a computational ontology based on healthcare security and privacy breaches, accompanied by a semantic similarity metric, to calculate to what extent the regulatory text covers what happened in breaches. Crowdsourcing requirements from regulatory text is also a promising approach. Breaux and Schaub \cite{breaux14crowd} compared the effectiveness of untrained crowd workers and trained experts on a requirements extraction task from privacy policies, and found that crowdsourcing can help reduce the cost of extraction while preserving accuracy.

\section{Conclusions}
\label{sec-conclusion}

In this paper, we presented the first analysis of the 2020 UK ICO regulation for age appropriate design using three online video sharing or live streaming platforms, popular among children -- TikTok, Twitch, and YouTube Kids. Our findings indicate, on the one hand, that some criteria in the regulation do not result in clear requirements, therefore, assessment of compliance becomes subjective. On the other hand, we identify a number of concerns in the studied platforms that might hinder the online protection of children.

Future work includes (i) incorporating additional steps from DPIA, specifically Steps 5 and 6 regarding risk assessment and mitigation, (ii) development of an ontology for age appropriate design and associated similarity-based computation of to what extent the ICO requirements are covered by the platforms, and (iii) conducting a human participant study to understand how various features provided by the online platforms are utilised, e.g. which parental controls are preferred by parents.

\bibliographystyle{splncs04}
\bibliography{kids-protection}

\begin{thebibliography}{10}
\providecommand{\url}[1]{\texttt{#1}}
\providecommand{\urlprefix}{URL }
\providecommand{\doi}[1]{https://doi.org/#1}

\bibitem{apthorpe2019}
Apthorpe, N., Varghese, S., Feamster, N.: {Evaluating the Contextual Integrity
  of Privacy Regulation: Parents’ IoT Toy Privacy Norms Versus COPPA}. In:
  Proceedings of the $28^{th}$ USENIX Security Symposium. pp. 123--140. USENIX
  Association (2019)

\bibitem{barata2020}
Barata, J.: {Regulating content moderation in Europe beyond the AVMSD}.
  [Online]
  \url{https://blogs.lse.ac.uk/medialse/2020/02/25/regulating-content-moderation-in-europe-beyond-the-avmsd/},
  last accessed 13/06/2020. (2020)

\bibitem{Breaux+Anton-08:regulatory}
Breaux, T.D., Ant{\'{o}}n, A.I.: Analyzing regulatory rules for privacy and
  security requirements. IEEE Transactions on Software Engineering
  \textbf{34}(1),  5--20 (Jan 2008)

\bibitem{breaux14crowd}
Breaux, T.D., Schaub, F.: Scaling requirements extraction to the crowd:
  Experiments with privacy policies. In: Proceedings of the 22nd International
  Requirements Engineering Conference (RE). pp. 163--172 (2014)

\bibitem{dpa2018}
{DPA}: {Data Protection Act 2018} (2018),
  \url{http://www.legislation.gov.uk/ukpga/2018/12/contents/enacted}

\bibitem{epic}
{Electronic Privacy Information Center (EPIC)}: {COPPA's Provisions}. [Online]
  \url{https://epic.org/privacy/kids/#Act}, last accessed 13/06/2020.

\bibitem{AVMSD}
{European Union}: {Audiovisual Media Services Directive} (2018),
  \url{http://data.consilium.europa.eu/doc/document/PE-33-2018-INIT/en/pdf}

\bibitem{coppa}
{Federal Trade Commission}: {Children’s Online Privacy Protection Rule}
  (2013),
  \url{https://www.ftc.gov/enforcement/rules/rulemaking-regulatory-reform-proceedings/childrens-online-privacy-protection-rule}

\bibitem{GDPR2018}
{GDPR}: {General Data Protection Regulation} (2018),
  \url{https://ico.org.uk/for-organisations/guide-to-data-protection/guide-to-the-general-data-protection-regulation-gdpr/}

\bibitem{Hashmi2015}
Hashmi, M.: A methodology for extracting legal norms from regulatory documents.
  In: Proceedings of IEEE 19th International Enterprise Distributed Object
  Computing Workshop. pp. 41--50 (Sept 2015)

\bibitem{hipaa03privacy}
HHS: Summary of the {HIPAA} privacy rule (2003), {United States Department of
  Health and Human Services (HHS)}.
  \url{http://www.hhs.gov/ocr/privacy/hipaa/understanding/summary/}

\bibitem{ico2020}
{ICO (Information Commissioner's Office)}: {Age appropriate design: a code of
  practice for online services}. [Online]
  \url{https://ico.org.uk/for-organisations/guide-to-data-protection/key-data-protection-themes/age-appropriate-design-a-code-of-practice-for-online-services/},
  last accessed 02/06/2020 (January 2020)

\bibitem{Kafali-17:Semaver}
Kafal{\i}, {\"O}., Jones, J., Petruso, M., Williams, L., Singh, M.P.: How good
  is a security policy against real breaches? {A HIPAA} case study. In:
  Proceedings of the 39th International Conference on Software Engineering
  (ICSE). pp. 530--540. IEEE Computer Society, Buenos Aires (May 2017)

\bibitem{kloess2019}
Kloess, J.A., Hamilton-Giachritsis, C.E., Beech, A.R.: {Offense Processes of
  Online Sexual Grooming and Abuse of Children Via Internet Communication
  Platforms}. Sex Abuse  \textbf{31}(1),  73--96 (2019).
  \doi{10.1177/1079063217720927}

\bibitem{lse-hate-2020}
Machackova, H., Blaya, C., Bedrosova, M., Smahel, D., Staksrud, E.:
  {Children’s experiences with cyberhate}. Tech. rep., London School of
  Economics and Political Science, EU Kinds Online. (2020),
  \url{https://doi.org/10.21953/lse.zenkg9xw6pua}

\bibitem{massey2010}
Massey, A.K., Otto, P.N., Hayward, L.J., Ant{\'o}n, A.I.: {Evaluating Existing
  Security and Privacy Requirements for Legal Compliance}. Requirements
  Engineering  \textbf{15},  119–--137 (2010).
  \doi{10.1007/s00766-009-0089-5}

\bibitem{nelson2016}
Nelson, B.: {Children's Connected Toys: Data Security and Privacy Concerns}
  (December 2016), \url{https://www.hsdl.org/?view&did=797394}, {Published by
  the United States Homeland Security Digital Library.}

\bibitem{nouwens2020}
Nouwens, M., Liccardi, I., Veale, M., Karger, D., Kagal, L.: {Dark Patterns
  after the GDPR: Scraping Consent Pop-ups and Demonstrating their Influence}.
  In: CHI '20: Proceedings of the 2020 CHI Conference on Human Factors in
  Computing Systems. p. 1–13. ACM (2020). \doi{10.1145/3313831.3376321}

\bibitem{ofcom2020-attitude}
{Ofcom}: {Children and parents: Media use and attitudes report 2019}. [Online]
  \url{https://www.ofcom.org.uk/\_\_data/assets/pdf\_file/0023/190616/children-media-use-attitudes-2019-report.pdf},
  last accessed 02/06/2020 (February 2020)

\bibitem{ofcom2020-attitude-annex}
{Ofcom}: {Children and parents: Media use and attitudes report 2019 (Annex)}.
  [Online]
  \url{https://www.ofcom.org.uk/\_\_data/assets/pdf\_file/0028/190576/children-media-use-attitudes-2019-annex.pdf},
  last accessed 02/06/2020 (February 2020)

\bibitem{reyes2018}
Reyes, I., Wijesekera, P., Reardon, J., On, A.E.B., Razaghpanah, A.,
  Vallina-Rodriguez, N., Egelman, S.: {Won’t somebody think of the
  children?” Examining COPPA compliance at scale}. In: Proceedings on Privacy
  Enhancing Technologies. pp. 63--–83 (2018). \doi{10.1515/popets-2018-0021}

\bibitem{riaz14re}
Riaz, M., King, J., Slankas, J., Williams, L.: Hidden in plain sight:
  Automatically identifying security requirements from natural language
  artifacts. In: Proceedings of the 22nd IEEE International Requirements
  Engineering Conference (RE). pp. 183--192 (2014)

\bibitem{sasha2019}
Shasha, S., Mahmoud, M., Mannan, M., Youssef, A.: {Playing With Danger: A
  Taxonomy and Evaluation of Threats to Smart Toys}. IEEE Internet of Things
  \textbf{6}(2),  2986--3002 (2019)

\bibitem{lse-survey-2020}
Smahel, D., Machackova, H., Mascheroni, G., Dedkova, L., Staksrud, E.,
  Ólafsson, K., Livingstone, S.and~Hasebrink, U.: {EU Kids Online 2020: Survey
  results from 19 countries}. Tech. rep., London School of Economics and
  Political Science (2020), \url{http://dx.doi.org/10.21953/lse.47fdeqj01ofo}

\bibitem{tiktok-duet}
{TikTok -- Privacy Controls}: [Online]
  \url{https://support.tiktok.com/en/privacy-safety/comment-duet-and-direct-message-control-default},
  accessed 06/07/2020

\bibitem{tiktok-terms-of-use}
{TikTok -- Terms of Service}: [Online]
  \url{https://www.tiktok.com/legal/terms-of-use}, accessed 22/06/2020

\bibitem{tiktok-community}
{TikTok Community Guidelines}: [Online]
  \url{https://www.tiktok.com/community-guidelines?lang=en, accessed
  22/06/2020}

\bibitem{tiktok-priv-policy}
{TikTok Privacy Policy -- EEA}: [Online]
  \url{https://www.tiktok.com/legal/privacy-policy?lang=en#privacy-eea},
  accessed 22/06/2020

\bibitem{twitch-community}
{Twitch -- Community Guidelines}: [Online]
  \url{https://www.twitch.tv/p/en-gb/legal/community-guidelines/}, accessed
  30/06/2020

\bibitem{twitch-cookies}
{Twitch -- Cookie Policy}: [Online]
  \url{https://www.twitch.tv/p/en-gb/legal/cookie-policy/}, accessed 30/06/2020

\bibitem{twitch-nudity-policy}
{Twitch -- Nudity, Pornography, and Other Sexual Content}: [Online]
  \url{https://www.twitch.tv/p/legal/community-guidelines/sexualcontent/},
  accessed 30/06/2020

\bibitem{twitch-priv-choices}
{Twitch -- Privacy Choices}: [Online]
  \url{https://www.twitch.tv/p/en-gb/legal/privacy-choices/}, accessed
  30/06/2020

\bibitem{twitch-priv-policy}
{Twitch -- Privacy Notice}: [Online]
  \url{https://www.twitch.tv/p/en-gb/legal/privacy-notice/}, accessed
  30/06/2020

\bibitem{twitch-moderation}
{Twitch - AUtoMod}: [Online]
  \url{https://help.twitch.tv/s/article/how-to-use-automod?language=en_US},
  accessed 30/06/2020

\bibitem{twitch-loot}
{Twitch Prime Loot Gifting}: [Online]
  \url{https://help.twitch.tv/s/article/twitch-prime-loot-gifting?language=en},
  accessed 30/06/2020

\bibitem{utz2019}
Utz, C., Degeling, M., Fahl, S., Schaub, F., Holz, T.: {(Un)informed Consent:
  Studying GDPR Consent Notices in the Field}. In: CCS '19: Proceedings of the
  2019 ACM SIGSAC Conference on Computer and Communications Security. p.
  973–990. ACM (2019). \doi{10.1145/3319535.3354212}

\bibitem{valente2017}
Valente, J., Cardenas, A.A.: {Security \& Privacy in Smart Toys}. In: IoT
  S\&P'17: Proceedings of the 2017 Workshop on Internet of Things Security and
  Privacy. pp. 19--24. ACM Press (2017). \doi{10.1145/3139937.3139947}

\bibitem{youtube-child-safety-1}
{YouTube -- Child Safety on YouTube}: [Online]
  \url{https://support.google.com/youtube/answer/2801999?hl=en-GB}, accessed
  25/06/2020

\bibitem{youtube-child-safety-2}
{YouTube -- Child Safety on YouTube}: [Online]
  \url{https://support.google.com/youtube/thread/12506319?hl=en}, accessed
  25/06/2020

\bibitem{youtube-community}
{YouTube -- Community Guidelines}: [Online]
  \url{https://www.youtube.com/intl/en-GB/about/policies/#community-guidelines},
  accessed 25/06/2020

\bibitem{youtube-create-content}
{YouTube -- Creating Content for YouTube Kids}: [Online]
  \url{https://www.youtube.com/yt/family/}, accessed 25/06/2020

\bibitem{youtube-kids-parental-guide}
{YouTube Kids -- Parental Guide}: [Online]
  \url{https://support.google.com/youtubekids/\#topic=6130504}, accessed
  25/06/2020

\bibitem{youtube-kids-notice-children}
{YouTube Kids -- Privacy Notice (for children)}: [Online]
  \url{https://kids.youtube.com/t/noticeforchildren}, accessed 25/06/2020

\bibitem{youtube-kids-priv-policy}
{YouTube Kids -- Privacy Notice (for parents)}: [Online]
  \url{https://kids.youtube.com/t/privacynotice}, accessed 25/06/2020

\bibitem{youtubekids-terms}
{YouTube Kids -- Terms of Service}: [Online]
  \url{https://kids.youtube.com/t/terms}, accessed 25/06/2020

\end{thebibliography}

\end{document}